\documentclass{moriond}
\usepackage{amsmath,amssymb}

\def\Journal#1#2#3#4{{#1} {\bf #2}, #3 (#4)}

\def\PRD{{\em Phys. Rev.} D}

\def\be{\begin{equation}}
\def\ee{\end{equation}}
\def\bea{\begin{eqnarray}}
\def\eea{\end{eqnarray}}

\newcommand\D{\mathcal D}
\def\L{\mathcal{L}}
\def\Lm{\mathcal{L}_m}

\newcommand{\Photo}{}

\begin{document}
\vspace*{4cm}
\title{Cosmology in Entangled Relativity}

\author{Olivier Minazzoli}

\address{Artemis, Universit\'e C\^ote d'Azur, CNRS, Observatoire C\^ote d'Azur\\ BP4229, 06304, Nice Cedex 4, France,\\Bureau des Affaires Spatiales, 2 rue du Gabian, 98000  Monaco}

\maketitle\abstracts{
General Relativity, in the absence of a cosmological constant, is an inevitable limit of Entangled Relativity, particularly when the universe is dominated by dust and/or electromagnetic radiation. In this communication, I emphasize that this arises from a specific type of decoupling termed \textit{intrinsic decoupling}. I then discuss what this implies for Dark Energy candidates within this framework. Furthermore, I introduce a novel and tantalizing hypothesis that the Lagrangian of Entangled Relativity represents merely the unperturbed term of an infinite series in a perturbative scheme. The terms of this series are dictated by the only dimensionful universal parameter of the theory, and notably, this series retains the \textit{intrinsic decoupling} of the original theory, non-perturbatively.
}

\section{Introduction}

Entangled Relativity is a somewhat novel general theory of relativity that is more economical than Einstein's General Relativity \cite{ludwig:2015pl,arruga:2021pr,minazzoli:2023mr}. Its name is not a reference to the phenomenon of quantum entanglement, but refers to the fact that matter and curvature are intertwined at the level of the formulation of the theory: one cannot formulate the theory of relativity without matter \cite{arruga:2021pr}. Otherwise stated, pure gravity is not possible within this framework, even in theory. Entangled Relativity was discovered by accident while trying to determine whether an $f(R,\Lm)$ theory could possess the \textit{intrinsic decoupling} observed in scalar-tensor theories with a non-minimal, but universal, multiplicative coupling between the scalar gravitational field and matter fields \cite{minazzoli:2013pr}. Although it might not be the only answer, the general structure of $f(R,\Lm)$ theories suggests that it is very unlikely to find other types of function $f$ that would also lead to the \textit{intrinsic decoupling}. Without this decoupling, the non-linear coupling between matter and curvature in the action would lead to significant deviations from General Relativity, which, conversely, explains many phenomena with a very high degree of accuracy.

As a result, one may describe Entangled Relativity as a \textit{top-down} approach, in that it was not specifically designed to address any particular issue within the standard physics paradigm, but rather, it emerged incidentally as a potential alternative to General Relativity.

The primary motivation to study Entangled Relativity is that it requires fewer universal dimensionful parameters than General Relativity to be defined, even within the framework of a path integral. A second key motivation relates to Einstein’s \textit{principle of relativity of inertia}, also known as \textit{Mach’s principle}. This principle, famously articulated by Einstein, posits that spacetime is entirely determined by matter, implying that vacuum solutions should not exist in a truly \textit{satisfying} theory of relativity \cite{einstein:1918an,einstein:1918sp}. Lastly, Entangled Relativity encompasses General Relativity as a limit in fairly generic situations, due to the \textit{intrinsic decoupling} mentioned previously. In a path-integral framework, the theory is formulated as follows \cite{minazzoli:2023mr} :
\be
Z = \int \D g  \prod_i \D f_i \exp \left[-\frac{i}{2 \epsilon^2} \int d^4_g x \frac{\L^2_m(f,g)}{R(g)} \right], \label{eq:ERPI}
\ee
where $\int \D$ relates to the sum over all possible field configurations, 
$R$ is the usual Ricci scalar that is constructed upon the metric tensor $g$, $ \mathrm{d}^4_g x := \sqrt{-|g|}  \mathrm{d}^4 x$ is the spacetime volume element, with $|g|$ the metric $g$ determinant, and $\L_m$ is the Lagrangian density of matter fields $f$---which could be the current \textit{standard model of particle physics} Lagrangian density, but most likely a completion of it. One can check that the dimension of what is historically called ``the action'' turns out to be an energy squared. Thus, the only undetermined dimensionful universal constant of the theory is a quantum of energy squarred
, denoted as $\epsilon^2$. This parameter is quantum in nature because it does not appear in the classical field equations---similar to $\hbar$ in standard physics---but it does influence the overall weight of all quantum paths. Alongside the causal structure constant $c$, which is implicit in the spacetime volume element, these are the only two universal constants of the theory.

The classical phenomenology of the theory, including the classical phenomenology of matter fields---which is defined from paths with stationary phases---can entirely be recovered by the following alternative action \cite{ludwig:2015pl}
\be
S \propto \int d^4_g x \frac{1}{\kappa}\left(\frac{R(g)}{2 \kappa} + \L_m(f,g) \right), \label{eq:ERGR}
\ee
with $\kappa$ a dimensionful scalar-field---although one may want to use a somewhat more usual definition of a dimensionless scalar-field with $\varphi := \sqrt{\Phi} := \bar \kappa / \kappa$, with $\bar \kappa$ a dimensionful normalisation constant with the dimension of $G/c^4$. One obtains
\be
S \propto \int d^4_g x \left(\frac{\Phi  R(g)}{2 \bar \kappa} + \sqrt{\Phi }\L_m(f,g) \right), \label{eq:ERGR2}
\ee
which corresponds to the class of theories with an \textit{intrinsic decoupling} with $\omega=0$ \cite{minazzoli:2013pr}.
Essentially, the \textit{intrinsic decoupling} in the field equations implies that the variation of $\kappa$ is negligible most of the time compared to the variation of the spacetime metric, allowing one to recover the action of General Relativity minimally coupled to matter. The effectiveness of this decoupling has been confirmed in the Solar System \cite{minazzoli:2013pr}, and surprisingly, it remains quite effective even for neutron stars \cite{arruga:2021pr}.
The field equations derived from the stationary phases in Eq. (\ref{eq:ERPI})---or from Eqs. (\ref{eq:ERGR}) and (\ref{eq:ERGR2})---are
\be \label{eq:genEReq}
G_{\mu \nu} = \kappa T_{\mu \nu} + F_R^{-1} \left[\nabla_\mu \nabla_\nu - g_{\mu \nu} \Box \right] F_R,
\ee
with 
\bea
\kappa = - \frac{R}{\Lm},\qquad
F_R := \frac{\partial F}{\partial R}= \frac{1}{2 \epsilon^2} \frac{\Lm^2}{R^2} = \frac{1}{2 \epsilon^2 \kappa^2},\label{eq:F_Rkappa}
\eea
where one has defined 
\be
F(R,\Lm):=-\frac{1}{2 \epsilon^2}\frac{\Lm^2}{R},\qquad \textrm{and}\qquad T_{\mu \nu} :=-\frac{2}{\sqrt{-g}} \frac{\delta\left(\sqrt{-g} \mathcal{L}_{m}\right)}{\delta g^{\mu \nu}}.
\ee 
Moreover, as is typical in $f(R)$ theories, the trace of the metric field equation allows us to derive the differential equation for the additional scalar degree of freedom:
\be
3 \kappa^{2}\Box \kappa^{-2}=\kappa \left(T-\mathcal{L}_{m}\right). \label{eq:sceq}
\ee
The entire set of field equations ensures that $\kappa$ remains well-defined in the $\Lm \rightarrow 0$ limit, as exemplified when the charge of a black hole approaches zero in this theory \cite{minazzoli:2021ej}. (A new paper discussing the properties of (electrically and magnetically) charged black holes in Entangled Relativity, both with and without spin, is currently under preparation).
\section{Cosmology after the radiation dominated era}

Assuming a Friedmann–Lemaître–Robertson–Walker metric, the equation on the scalar degree-of-freedom reduces to
\be \label{eq:phi}
\ddot F_R + 3 H \dot F_R = \kappa \left(\mathcal{L}_{m}-T\right) ,
\ee
where $H:=\dot a / a$ is the Hubble parameter. For dust, as well as for electromagnetic radiation, one has $\Lm = T$, such that Eq. (\ref{eq:phi}) reduces to $\ddot F_R + 3 H \dot F_R = 0$,
and the friction term from the cosmological expansion at the begining of the matter dominated era freezes the scalar degree-of-freedom $\dot F_R \propto a^{-3}$.
It is then straightforward to verify from Eq. (\ref{eq:genEReq}) that one asymptotically recovers General Relativity without a cosmological constant. Consequently, General Relativity without a cosmological constant becomes an unavoidable limit of Entangled Relativity, provided the material content of the universe satisfies $\Lm = T$.

In our current universe, the condition $\Lm \approx T$ holds, rather than $\Lm = T$. For instance, cosmological magnetic fields result in $\Lm \propto B^2 \Rightarrow \Lm \neq T$, but their amplitude is expected to be negligible compared to other material contents. Consequently, careful analyses are required to determine how even the late-time cosmological behavior might slightly diverge from the predictions of General Relativity. 

Hence, it appears that in Entangled Relativity, the acceleration of the universe's expansion must stem from some form of Dark Energy, rather than from a cosmological constant. However, it is important to note that any Dark Energy candidate that does not satisfy $\Lm = T$---such as a simple quintessence field that is such that $\Lm = P$---would not result in a decoupling, preventing the theory from reducing to General Relativity with Dark Energy. More specifically, such a scenario would predict variations in Newton's constant $G$ (since $G \propto \kappa$), against which there are relatively stringent constraints.
\section{Cosmology during the radiation dominated era and before}

In order to study what happens in detail during the matter-dominated cosmological era, one must first understand well enough what occurred before, particularly to determine what kind of initial conditions one may expect for the scalar degree-of-freedom at the onset of the matter-dominated era. This is currently uncharted territory, but I would like to stress that any such study would first necessitate deriving the on-shell matter Lagrangian from first principles for the various complicated states of matter during the radiation era. For instance, it would be necessary to derive the on-shell matter Lagrangian for a quark-gluon plasma in Eq. (\ref{eq:phi}).

However, given that Entangled Relativity likely deviates from General Relativity under the conditions of the early universe, there may be room for the theory to address some of the current tensions observed with the $\Lambda$CDM model.

\section{Perturbative Entangled Relativity}

Let us assume that Entangled Relativity represents the leading term in an infinite expansion based on the theory's sole dimensionful universal constant, $\epsilon^2$. This kind of expansion is anticipated if Entangled Relativity forms the leading term of a perturbative expansion of another fundamental theory, which would depend on only two dimensionful universal constants: $\epsilon^2$ and $c$. Such an expansion would read as follows:
\be  \label{eq:EERp}
F(R,\Lm) = \sum_{n=1}^{\infty} \frac{1}{\epsilon^{2n}}\frac{\omega_n}{2n} \frac{\Lm^{2n}(f,g)}{R^{3n-2}(g)},
\ee
where $\omega_n$ are adimensional constants and $\omega_1 = -1$ in order to recover Entangled Relativity at leading order. 
The form of the non-linear couplings in the series is dictated by their dimensionality. The metric field equation can be casted under the following form
\be \label{eq:eerMeq}
G_{\mu \nu} + \Lambda g_{\mu \nu} = \kappa T_{\mu \nu} + F_R^{-1} \left[\nabla_\mu \nabla_\nu - g_{\mu \nu} \Box \right] F_R,
\ee
where $\Lambda$, $\kappa$, and $F_R$ are extremely complicated terms defined by the combination of different infinite series of non-linear $f_n(R, \Lm)$ couplings. As with previous examples, the differential equation for the scalar degree-of-freedom emerges after taking the trace of the metric field equation, and it reads as follows:
\be \label{eq:scalarGen}
3 F_R^{-1} \Box F_R = \kappa T + R - 4 \Lambda.
\ee
One can demonstrate that, despite $\kappa$ and $\Lambda$ consisting of extremely complicated infinite series of non-linear $f_n(R,\Lm)$ couplings, the overall right-hand side of Eq. (\ref{eq:scalarGen}) simplifies non-perturbatively to yield:
\be
3 F_R^{-1} \Box F_R = \kappa (T-\Lm). \label{eq:Sdof_dec}
\ee
As a consequence, in a universe dominated by dust, where $\Lm = T$ roughly as in our current universe, the scalar degree-of-freedom inevitably freezes ($\dot F_R \propto a^{-3}$), leading to the asymptotic metric field equation:
\begin{equation}
G_{\mu \nu} + \Lambda_a g_{\mu \nu} = \kappa_a T_{\mu \nu},
\end{equation}
where $\Lambda_a$ and $\kappa_a$ represent the asymptotic values of $\Lambda$ and $\kappa$ over cosmic time. This equation resembles General Relativity with a cosmological constant, suggesting that it is an inevitable non-perturbative limit (for $\Lm=T$) of a theory whose perturbative expansion takes the form of Eq. (\ref{eq:EERp}). Such a theory is unlikely to be String Theory, as although perturbative String Theory also relies on only two dimensionful constants,\cite{Veneziano} one of them is a fundamental length (the string length $\ell_s$) rather than an energy like $\epsilon$. Additionally, unlike standard physics with the dimensionful universal constants $\hbar$, $G$, and $c$, a fundamental length cannot be converted into a fundamental energy, and vice versa, if the only remaining dimensionful constant is the speed $c$.

\section{Conclusion}

Entangled Relativity represents an entirely new direction to explore and evaluate. This communication serves as yet another motivation to encourage the community to delve into it and to discover what can realistically be expected from it, particularly at the cosmological level.


\section*{References}

\begin{thebibliography}{99}
\bibitem{ludwig:2015pl} H. Ludwig, O. Minazzoli and S. Capozziello \Journal{Physics Letters B}{751}{576}{2015}
\bibitem{arruga:2021pr} D. Arruga, O. Rousselle and O. Minazzoli, \Journal{\PRD}{103}{024034}{2021}
\bibitem{minazzoli:2023mr} O. Minazzoli, \Journal{Proceedings of the Gravitation session of the 57th Rencontres de Moriond}{}{}{2023}
\bibitem{minazzoli:2013pr} O. Minazzoli and A. Hees, \Journal{\PRD}{88}{041504}{2013}
\bibitem{einstein:1918an} A. Einstein, \Journal{Annalen der Physik}{360}{241--244}{1918}
\bibitem{einstein:1918sp} A. Einstein, \Journal{Sitzungsberichte der K{\"o}niglich Preu{\ss}ischen Akademie der
  Wissenschaften}{}{270-272}{1918}
\bibitem{minazzoli:2021ej} O. Minazzoli and E. Santos, \Journal{European Physical Journal C}{81}{640}{2021}
\bibitem{Veneziano} G. Veneziano, \Journal{Europhys. Lett.}{2}{199}{1986}


%
%
%

\end{thebibliography}
\end{document}